# Analysis of the Christensen *et al.* Clauser-Horne (CH)-Inequality-Based Test of Local Realism

Donald A. Graft

donald.graft@cantab.net

## ABSTRACT

The Clauser-Horne (CH) inequality [1] can validly test aspects of locality when properly applied [2]. This paper analyzes a recent CH-based EPRB experiment, the Christensen *et al.* experiment [3]. Full details of the data analysis applied to the experiment are given. An alternative analysis is also presented that considers the role of accidental coincidences and confirms and justifies the main analysis. It is shown that the experiment confirms locality and disconfirms the quantum joint prediction. To make sense of this surprising finding, the conclusion presents a new 'rational interpretation' of the EPR paradox. The paper also contributes to promulgation of robust and correct data analysis by describing the important degrees of freedom that affect the analysis, and that must be addressed in the analysis of any EPRB experiment.

## 1. INTRODUCTION

In a previous paper [2] I showed that the Clauser-Horne (CH) inequality [1] can validly test aspects of locality when properly applied. However, the considerations adduced there do not exhaust the methodological concerns that apply to an actual CH experiment and its interpretation. The experiment produces only cold, raw data. A complicated process remains to move from the raw data to a statement about violation (or not) of the CH inequality. In published papers we typically do not see full accounts of the data analysis process, and we are expected to trust the authors' analyses. The Christensen *et al.* experiment [3] and data analysis present a perfect case in point of the need to critically consider proffered data analyses. It is shown that the analysis relies upon unjustified post-selection (data discarding) that produces an artifactual violation of the CH inequality. This faulty analysis provides the motivation for the correct analysis presented in this paper. A decisive result about locality is so full of implications for the foundations of physics that it cannot be based on an analysis whose specifics are hidden in the shadows. All experimental reports should include (or link to) full details and source codes of the data extraction/analysis process, starting with the raw data and ending with a statement about confirmation or disconfirmation. The raw experimental data itself should also be made readily available without restriction to all researchers for independent analysis of the data. Christensen *et al.* have cooperated generously in meeting these criteria and I applaud them.

This paper analyzes a recent CH-based EPRB experiment, the Christensen *et al.* experiment [3]. I give full details of the data analysis applied to the experiment and show that the analysis confirms locality and disconfirms the quantum joint prediction. The paper is organized as follows. In Section 2, the Christensen *et al.* data analysis is considered and found to be faulty, thereby motivating an independent, correct analysis. In Section 3, the methodology to be applied in the analysis of this paper is developed. This section, while developing a sound foundation for the analysis, also contributes to promulgation of robust and correct data analysis by describing the important degrees of freedom that affect the analysis, and that must be addressed in the analysis of any EPRB experiment. In Section 4, the quantum joint prediction for the conditions of the Christensen *et al.* experiment is obtained for comparison to the experimental results. In Section 5, the experimental data is analyzed according to the developed methodology. In Section 6, an alternative analysis of the data is developed to respond to possible objections to the analysis of Section 5. In Section 7, the results of the independent analysis are discussed and conclusions are drawn.

## 2. CRITIQUE OF THE DATA ANALYSIS PERFORMED BY CHRISTENSEN *ET AL.*

Christensen *et al.* [3] claim that their experiment violates the CH inequality and therefore confirms the quantum joint prediction (thus also confirming quantum nonlocality), but they have not formally published details of the methodology and analysis that they used to move from the raw data to their claim. Naturally, one is curious to know if any significant errors or additional (false) assumptions beyond the bare CH inequality conditions are present in the Christensen *et al.* data analysis. Upon an inquiry, Christensen kindly provided me with MATLAB code that generates the published counts and CH metrics found in [3]. I have cleaned up the code by removing unrelated things and renaming some variables to make the code more readable and understandable. The revised code generates the published counts and fully embodies the algorithm used by Christensen *et al.* to analyze the data. The code is available as 'ChristensenAnalysis.m' online [6]. To run this code, place all the original MATLAB data files (not the extracted data files used in my analysis) into a single directory and replace the directory path in the code with the chosen directory path.

Perusal of the code immediately reveals several serious errors that disqualify it as a valid analysis of the experimental data. Most crucially, inspection of the algorithm reveals that the main loop runs once for each Pockels cell opening trial, and that for each run of the loop, only one detection event (singles count) can be registered for each side, and thus only one coincidence count can be registered. Technically, the loop is run twice to account for the missing Pockels cell opening (to be described later in this paper), but for each run of the loop this limitation applies, i.e., overall only one detection event per side can be registered for any given trial. While this limitation would not be a problem if there were a maximum of only one source event per trial, analysis of the event histograms (Section 3.8) shows that there is in fact a Poissonian distribution of detections per trial. Therefore, this limitation of the analysis leads to a serious and unjustified post-selection of the data (data discarding) that produces a false, artifactual violation of the CH inequality. To leave no doubt that this post-selection is solely responsible for the claimed violation I created a fixed version of the Christensen *et al.* analysis code, which correctly includes all of the experimental data. This fixed code is available as 'ChristensenAnalysisFixed.m' online [6]. Following is the output of the *unfixed* code 'ChristensenAnalysis.m':

```
>> [cou,sbell] = generateQRNGdata
Elapsed time is 38.075708 seconds.

cou =

        46068       29173       46039     27153020
        48076       34145      146205     28352350
       150840       34473       47447     27827318
       150505        1862      144070     27926994

sbell =

  5.8701e-005
```

The counts are the same as the published counts, however, the calculated CH metric (`sbell`) is a little higher than the published value because here singles-rate averaging (singles rate averaging is discussed in Section 3.5) is not used. It can be seen that CH appears to be violated, because `sbell` is positive. Following is the output of the *fixed* code 'ChristensenAnalysisFixed.m':

```
>> [cou,sbell] = generateQRNGdataFixed

Elapsed time is 38.928476 seconds.

cou =

       46960        29221        46971     27153020
       49048        34203       148026     28352350
      153728        34513        48100     27827318
      153531         1868       146103     27926994

sbell =

 -3.4379e-006
```

All of the counts are somewhat higher here because the excluded events are now included. Most importantly, the CH metric has become negative, indicating no violation of the CH inequality. It is thus clear that the reported violation is due to post-selection of the experimental data (use of singles-rate averaging does not change this conclusion).

This discrepancy between the results of full counting versus post-selected counting, an experimental finding that cannot be denied, cries out for explanation and we are driven to speculate on the mechanisms that may be producing the effect. An obvious possibility is that the post-selection leads to unfair sampling. The post-selection can be expected to affect singles counts more than coincidences, and because the singles counts appear with negative weights in the CH inequality, the post-selection unfairly biases the CH metric in favor of a violation. For example, suppose a trial contains two detections at side A and one detection at side B. The analysis code registers only one detection for each side and one coincidence. The additional single count is not registered. Christensen *et al.* nowhere describe and justify this post-selection, either in the published paper [3] or its supplementary material. I believe that the justification is based upon an erroneous belief that this counting method eliminates any possible 'coincidence loophole' effects. Further consideration thereof is beyond the scope of this paper, though I remark that if we do not appeal to a coincidence effect there is no need to decimate the data in an effort to eliminate said effect. This paper aims primarily to report the discovery of the intriguing anomaly in the Christensen *et al.* data analysis. My own analysis described later in this paper properly includes all of the experimental data, and this difference alone is sufficient to account for the contradiction between the two analyses. For a valid and robust conclusion, *data must not be post-selected*. Section 6 further discusses the full-counting method and its justification.

Another possible mechanism to account for the effect relies upon the finite convergence rate of the CH metric as the experiment progresses. When one post-selects, it is as if the experiment has a smaller number $N$ of source events, so the experiment may converge at a slower rate compared to full counting. If the experiment is then stopped before full convergence, which is usually the case as the experiments tend to be relatively short (i.e., nowhere near the infinite $N$ required to fully converge), the two counting measures will necessarily differ. If one method converges from above to a CH metric below 0, while the post-selected method lags behind, the effect can be easily accounted for. Unfair sampling and convergence effects may both be involved, and future work is required to assess the extent of each.

This serious error of the analysis, post-selection, which alone accounts for the artifactual violation of the CH inequality, seems to render superfluous consideration of any further errors; nevertheless, several more should be exposed in the interest of promulgating robust analytical methods. Thus continuing, an inefficient implementation of the analysis code forces the use of heuristics that further post-select the data. The code runs a loop over the trials and then within the loop runs another loop looking for detection events. Algorithmically, the analysis is $O(T^2)$, where T is the total number of trials. The analysis of this paper is $O(T)$, which allows for highly efficient analysis. Due to the gross inefficiency, coupled with the slowness of MATLAB interpreted code, Christensen *et al.* implemented an arbitrary heuristic to accelerate the algorithm, which otherwise would have been intolerably slow. The heuristic is implemented in the code by a variable *searchrange*, which is set to a value of 5 to generate the published counts. Problematically, changing the value of the arbitrary variable *searchrange* changes the counts produced by the algorithm. For example, changing *searchrange* to 10 increases the counts and we see that this heuristic post-selects the data. This mechanism is less serious than the erroneous counting previously described, but again, for a valid and robust conclusion, data must not be post-selected.

Mindful that the aforementioned errors are already disqualifying, one additional error should be reported. The data as published, due to technical limitations of the experiment, is missing half of the Pockels cell opening events and the authors specify that these events must be added back during analysis. However, Christensen *et al*. add back these events too late in the analysis, running separate analytical loops over two disjoints sets of events, thereby undermining the event matching. The way to handle this problem correctly consists of adding the missing openings explicitly in the raw data, re-ordering the events into time sequence, and then proceeding with the main analysis. That is the strategy adopted in this paper's analysis.

The described errors show the Christensen *et al*. analysis to be invalid and therefore unacceptable. We are motivated to conduct an independent, correct data analysis that avoids the described analytical errors.

## 3. METHODOLOGY OF THE EXPERIMENTAL DATA ANALYSIS

Following the methodology developed and justified in [2], the CH inequality is applied to the output of simulations and the Christensen *et al*. experimental data [4]. The metrics considered are the raw CH metric in both linear and ratio forms and the positivity (fraction of runs that violate CH). The quantum joint prediction is obtained from simulation and compared to the experimental data, resulting in confirmation or disconfirmation of the quantum joint prediction (which may bear on locality in appropriate physical arrangements).

Proceeding from the raw data to a conclusion confirming or disconfirming the quantum joint prediction involves many steps and degrees of freedom. We need to understand these and make good decisions on how to proceed so that our conclusions are robust and reliable. The following degrees of freedom are relevant to the arguments of this paper, and they should be taken into account in any analysis of simulations and experimental data.

### 3.1 Source state

The source events are considered to be nonmaximal singlets with arbitrary maximality. Nonmaximal states are required to show definitive violation of CH at current detection efficiencies [5]. Following Eberhard and Christensen *et al*., I denote the degree of maximality by the factor $r$. Christensen *et al*. report using a nonmaximal singlet source with $r = 0.26$. To make our quantum joint prediction, we therefore also use $r = 0.26$ in simulations. Eberhard, in his pioneering work [5], used $r = 0.311$ (at 75% efficiency). Due to the lower maximality we expect the Christensen *et al*. experiment to be less sensitive than allowed for by the Eberhard analysis, and we have to verify that the reported efficiency of 75% is definitive at $r = 0.26$.

### 3.2 Measurement angles

For analysis of the experimental data, we must accept the experimentally arranged angles, and these angles do not enter into consideration for the analysis. For simulations, however, the angles are a degree of freedom. For example, to obtain the quantum joint prediction for the CH metric at a given maximality and detector efficiency, a simulation can search the angle space for the highest metric value. An analytical solution giving these values is difficult and the angle space landscape is mysterious, so search of the multidimensional angle space is a pragmatic and effective strategy.

### 3.3 Detection efficiency

Detectors do not register all source events; detections are missed. If both channels always miss the same events, it is just as if a shorter overall experimental run occurred, and there is no consequence for analysis of simulations and experiments (with sufficiently long experiments). If the detection losses are independently distributed at the two sides, however, detection efficiency has a large impact on the predicted and observable CH metrics. Christensen *et al*. report 75% efficiency for the detectors in their experiment. While Eberhard [5] shows that CH violations are predicted by the quantum joint prediction for 75% efficiency or better at maximality $r = 0.311$, the Christensen *et al*. source maximality is only $r = 0.26$, so it is important to reconsider the efficiency minimum in the experiment to be sure that a violation is theoretically possible. In simulations, we are free to set the detection efficiency and investigate how the efficiency affects the predicted metrics.

## 3.4 Partition size

Our CH positivity metric for a given overall run represents the fraction of contiguous sub-runs of length P (the partition size) that violate CH. For example, we may run a 10,000-event experiment 100 times. Here the partition size is 10,000, and the positivity is the number of violating sub-runs divided by 100. We would like to use a large partition size but simulations and searches must remain tractable, and the Christensen *et al.* experimental data is not long enough to allow for very large partitions. For analysis of the experiment and for simulations, the aforementioned partition size of 10,000 is convenient and sufficiently large for statistical accuracy. Nevertheless, the partition size is free in the simulations and the experimental data analysis, so investigation of the effect of varying partition sizes on the CH metrics is relevant and interesting.

## 3.5 Singles-rate averaging

The CH inequality contains a singles rate for each of the two sides. Technically, a single experimental arrangement is used, e.g., we use singles at A and singles at B in runs when the angle set is a1b1, and not for a1b2, a2b1, or a2b2 (the experimental run to use depends on which of the four CH inequality variants is used). Some workers, however, choose to average the singles rate of two experiments. For example, to obtain the singles rate for side A, the values seen in the a1b1 and a1b2 experiments are averaged. The intent presumably is to improve the statistics, but my simulations show that singles-rate averaging has a deleterious effect on the CH metrics. We will see that at $r = 0.26$ and *efficiency* $= 0.75$, the quantum joint prediction yields CH violation only when singles-rate averaging is not used. When singles-rate averaging is used, no violation is predicted. Nevertheless, singles-rate averaging is a free parameter for analysis of simulations and experimental data, so we will report results with and without singles-rate averaging. As the derivation of the CH inequality does not allow for singles-rate averaging and its use attaches unneeded additional (uncertain) hypotheses not clearly related to locality we should view it suspiciously.

## 3.6 Event pairing

In a simulation it is easy to pair detection events at the two sides and relate them to the source events. In experimental data, however, it is not so easy to identify valid detection event pairs due to the time-of-flight differential between the two sides, detection jitter, and the presence of detections due to noise and accidental coincidences. An explicit pairing method must be defined and documented. The usual strategy is to correct the timestamps of the detection events with estimated time-of-flight delays and then look for event pairs close enough together in time to have come from a single source event. For the Christensen *et al.* experiment, we also have the Pockels cell opening times available as a stable time reference. Therefore, we correct for time-of-flight and then look for detection event pairs where each detection time of the pair is within the same valid Pockels cell opening window.

The Christensen *et al.* experiment used a Pockels cell opening duration of 2 microseconds every 40 microseconds, so it would seem desirable to use a window size of 2 microseconds for event pairing. However, the TES detector has on the order of 100-500 nanoseconds of jitter and so we might like to increase the window size to allow for that. On the other hand, reducing the window size may improve things by excluding more noise. Therefore, the effect of the window size on the positivity was investigated, see Table 6. It is clear that the positivity is maximized at the expanded window size of 2.5 microseconds. There is also a peak of positivity at a window size of 1.25 microseconds, but this window size is still inferior to the maximum found at 2.5 microseconds. In a private communication, Christensen advised me to use the expanded window size, but in the published material, a window size of 1.2 microseconds is used. I have thoroughly explored the behaviors at different window sizes and I find no value of window size for which the 50% rule is significantly violated. To maximize the chances of finding a CH violation in the experimental data, therefore, I use a window size of 2.5 microseconds in the analysis to be reported. None of the conclusions of this paper are changed by instead using a window size of 1.2 or 2.0 microseconds.

## 3.7 Time-of-flight delay corrections

To allow for event pairing, we have to correct for the times of flight of source events to the detectors for each of the two sides, and align them to the Pockels cell openings. These delays are difficult to measure and so the delays are typically determined directly from the experimental data. For example, the 2-dimensional delay space can be searched for delay values that maximize the CH metric, or the number of coincidences, etc. The resulting delays are used for subsequent analysis. To adopt an analysis most favorable to finding CH violations, I choose to use the delays that maximize the CH metric. I also report on results using the delays reported by Christensen, *et al*.

### 3.8    Normalization and post-selection

The Christensen *et al.* experiment implements Pockels cell openings to create 'trials'. The Pockels cell is opened for 2 microseconds every 40 microseconds to generate trials. The count of these trials is used to normalize the detection counts, i.e., to convert them to probabilities for use in the CH inequality. This design implicitly supposes that there is only one source event per trial, but this is not satisfied in the real world, where there is a Poissonian distribution of source events per trial as well as background noise events. In a perfect world, we would never get more than two events per trial, corresponding to a single source event that may result in two detections events, one per side. The source intensity can be adjusted to affect the average number of events per trial, but a strict single-source-event-per-trial distribution is not obtainable. For example, considering the overall Christensen *et al.* data (using the 2.5 microsecond window previously described, no singles-rate averaging, and CH-optimal delay corrections), we obtain the following histogram for the number of detections per trial at side A:

```
Count of detections: number of trials with that count
0: 390174
1: 388931
2: 5766
3: 604
4: 155
5: 65
6: 27
7: 15
8: 10
9: 3
10: 1
11: 0
12: 1
13: 1
14: 0
15: 0
16: 0
```

The Poissonian character of the distribution is clearly evident and we therefore see that the intent of the experimental design (to have one source event per trial) is not reached and that the normalization may be questioned. For practical purposes of data analysis, we can either accept all the events regardless of their distribution in trials, or we may consider trials with more than some number of events to be pathological (possibly caused by source intensity fluctuations, noise bursts, or some other anomalies) and we might consider excluding the extra events. Obviously this degree of freedom of the analysis applies only to the experimental data, because in simulations we can easily enforce one source event per trial. As it is critical to include all the experimental data and to avoid any arbitrary post-selection, we do not exclude any events from the experimental analysis. We have seen that the Christensen *et al.* data analysis includes a maximum of only two events per trial (a single detection at each side), and that due to the failure of the one-source-event-per-trial assumption, the data is post-selected. This post-selection leads Christensen *et al.* to erroneously derive a violation of the CH inequality from the data set.

Above I noted that the normalization by trial counts in the Christensen *et al.* experiment can be questioned, because, instead of there being always exactly one source event per trial, a Poissonian-like distribution of events per trial with an expectation of well below 1.0 is seen in the experimental data. The distribution thus has a scaling effect on the CH metric through normalization that affects its absolute value. Use of the positivity metric, however, factors out the arbitrary absolute CH metric, and its use allows application of the 50% rule despite the existence of a Poissonian source rate distribution. The normalization appears valid as long as it is the same for all the measurement angle combinations, i.e., the terms of the CH inequality are not normalized at different scales. Even with this minimal interpretation of normalization, we see that normalization by trial counts *does* properly eliminate the effect of possible different numbers of trials for the different measurement angle combinations. Fully correct normalization requires knowledge of the exact numbers of source events in the four angle combinations. These counts are unknown in a real experiment, so any normalization applied cannot be fully correct. The view here is that normalization by trial counts is a useful and necessary operation, and that the absolute CH metric is factored out by considering positivity.

### 3.9    Background detections

In a real experiment detection events not associated with a source singlet event are encountered. These detections can be caused by ambient light, thermal noise, etc., collectively referred to as 'noise'. Like the experimental detection efficiency, we have no control over the noise, and it does not enter into the experimental data analysis. If the analysis shows no CH inequality violation, one could argue that it is due to too much noise, or too little efficiency, but one cannot claim confirmation of the quantum joint prediction, because numerous local models can generate similar metrics. One can conclude only that possibly the experiment was inadequate, or if it was indeed adequate, that the quantum joint prediction was disconfirmed.

We now turn to the quantum joint prediction for the Christensen *et al*. experiment and the experimental data analysis, following the methodology and degrees of freedom described above.

## 4.    THE QUANTUM JOINT PREDICTION

It is important to obtain the quantum joint prediction for the conditions of the Christensen *et al*. experiment for two reasons. First, we need to verify that the CH inequality can indeed be violated under those conditions; if it cannot, then our work is done and we conclude that the experiment cannot confirm or disconfirm the quantum joint prediction. Second, we can compare the experimental results with the predicted results to determine if the experiment tends to confirm or disconfirm the quantum joint prediction.

To obtain the quantum joint prediction a computer simulation was used (the source code for the simulation is available online [6]). A quantum joint PDF with $r = 0.26$ is randomly sampled. The efficiency, partition size, and noise level are configurable. The program operates as follows: First, one measurement angle is chosen as 0 radians (this can be done without loss of generality due to the rotational invariance of the singlet state) and the other three measurement angles are chosen randomly. Choosing one angle as 0 radians allows use of a 3-dimensional search of the angle space instead of a 4-dimensional search. Second, the simulation is run 100 times with the configured efficiency, partition size, and noise level and the CH metrics are calculated. Third, from the random starting angle set, the program performs a 3-dimensional optimization using Powell's direction set method to search the angle space to find the local maximum of the CH metric starting from the random starting point. The simulation repeats continually with new starting points, so that after a long run, the global maximum of the CH metric is found. As it is running, the program tracks the highest CH metric so far encountered. Fourth, each time a new highest CH metric is found, the program runs the (100 x partition size) simulation run again ten times and reports the average CH metrics across the ten runs. This allows for fast searching of the angle space as well as accurate metrics for the reported solutions. Finally, after the program is allowed to run for a lengthy time, the final CH metrics are reported.

Note that the simulation considers and reports solutions only for the fourth of the four variants of the CH inequality derived in [2] and as used by Christensen *et al*. This choice was made because the angle set used for the Christensen *et al*. experiment produces positivity greater than 0.0 only for the fourth variant of the inequality.

Table 1 shows the solutions for different efficiencies found for the following conditions: $r = 0.26$, partition size 10,000, and no background (noise) counts. It can be seen that, without singles-rate averaging, the CH inequality is significantly violated at 75% efficiency. However, with singles-rate averaging, the CH inequality is *not* violated at 75% efficiency, and it can be seen that the CH metrics are reduced by averaging at any efficiency level. Because Christensen *et al*. claim a 75% efficiency, it is therefore a mistake to analyze the experimental data with singles-rate averaging. Nevertheless, we keep an open mind and report the analysis of the data with and without averaging.

Table 2 shows the solutions found for different partition sizes for the following conditions: $r = 0.26$, efficiency 0.75, no background (noise) counts, no singles-rate averaging. It can be seen that only at a very small partition size (100) does the positivity drop below 0.5, indicating that the CH inequality is not violated. We prefer to have a large partition size to obtain a predicted positivity significantly greater than 0.5, but in practice the partition size must be limited to make simulations tractable and because the data from the experiment is limited. Using a partition size of 10,000 is a good compromise that still gives a high positivity. The experimental data can be analyzed with different partition sizes, however, and the behavior can be compared to Table 2.

I emphasize the usefulness of the positivity as a metric for application of the 50% rule [2]. The 50% rule tells us that we need to see statistically significant positivity greater than 0.5 to claim a violation of the CH inequality. The positivity metric eliminates the problem of arbitrarily diminished absolute CH metrics (due to source state nonmaximality, detection inefficiency, noise, etc.); and by means of this normalization, the different experiments and simulations become comparable.

Table 3 shows the solutions found for different noise levels for the following conditions: $r = 0.26$, efficiency 0.75, partition size 10,000, no singles-rate averaging. There are many ways to model noise in a simulation. My intent here is to demonstrate the degradation of the CH metric caused by noise. For the simulation, noise is implemented by generating additional detection events independently at each side with a probability given by the noise value shown in Table 3. For further details, refer to the simulation source code online [6]. It is clear from Table 3 that noise dramatically reduces the CH metric. We must assume that in the Christensen *et al.* experiment noise is very low, due not only to the experimental design (purely local without long fiber optics, good shielding, etc.), but also due to the event pairing with a small window size. Otherwise, the experiment cannot be decisive. Although not directly comparable statistically, the background level reported by Christensen *et al.* is consistent with a noise value of 0.001 or less in Table 3, and if that is the case, then the experiment can be decisive, either for or against the quantum joint prediction.

The results of this section lead us to expect that if the quantum joint prediction is valid in the experiment, then the analysis of the experimental data must produce values broadly in line with the values of Table 2. Certainly, we expect to see a CH violation with high positivity, but we also expect the positivity to vary with partition size as shown in the table. We turn now to analysis of the Christensen *et al.* experimental data.

**Table 1. Quantum joint prediction at various efficiencies**
**(solutions for *r* = 0.26, 100 runs of partition size 10,000,**
**no noise counts)**

| Efficiency | Without singles-rate averaging | | | With singles-rate averaging | | |
|---|---|---|---|---|---|---|
| | CH | CH ratio | Positivity | CH | CH ratio | Positivity |
| 0.65 | -0.009210 | 0.895301 | 0.004000 | -0.011290 | 0.853898 | 0.000000 |
| 0.70 | -0.003110 | 0.968371 | 0.200000 | -0.006507 | 0.923270 | 0.000000 |
| 0.75 | 0.003794 | 1.037920 | 0.848000 | -0.000975 | 0.989276 | 0.292000 |
| 0.80 | 0.011927 | 1.107756 | 1.000000 | 0.005585 | 1.056602 | 1.000000 |
| 0.85 | 0.020750 | 1.174020 | 1.000000 | 0.012717 | 1.119401 | 1.000000 |
| 0.90 | 0.031512 | 1.227730 | 1.000000 | 0.020821 | 1.185587 | 1.000000 |
| 0.95 | 0.042810 | 1.297279 | 1.000000 | 0.030012 | 1.251386 | 1.000000 |
| 1.00 | 0.056044 | 1.322048 | 1.000000 | 0.039829 | 1.315643 | 1.000000 |

**Table 2. Quantum joint prediction at various partition sizes**
**(solutions for *r* = 0.26, efficiency = 0.75,**
**no singles-rate averaging, no noise counts)**

| Partition size | Positivity |
|----------------|------------|
| 100 | 0.471000 |
| 500 | 0.573000 |
| 1000 | 0.604000 |
| 5000 | 0.767000 |
| 10,000 | 0.847000 |
| 25,000 | 0.957000 |
| 50,000 | 0.986000 |

**Table 3. Quantum joint prediction at various noise levels**
**(solutions for *r* = 0.26, efficiency = 0.75,**
**no singles-rate averaging, partition size = 10,000)**

| Noise | Positivity |
|-------|------------|
| 0.000 | 0.872000 |
| 0.001 | 0.669000 |
| 0.002 | 0.524000 |
| 0.003 | 0.229000 |
| 0.004 | 0.118000 |
| 0.005 | 0.052000 |

## 5. EXPERIMENTAL DATA ANALYSIS

The data analysis begins with the raw experimental data published by Christensen *et al*. [4]. A reliable and replicable data analysis must document all the steps that lead from the raw data to a binary decision: "the dataset supports the quantum joint prediction" or "the dataset does not support the prediction". Here I document all of these steps.

### 5.1    Data extraction

The Christensen *et al*. raw data is distributed by the authors as a directory tree of MATLAB files. The raw data as published is inconvenient for analysis for several reasons. First, the overall data set is contained in a directory tree of MATLAB files, with the angle settings encoded in the file names and the results (channel detections and timetags) stored in arrays in each MATLAB file. Extracting the data for analysis is nontrivial and error-prone. It would be desirable to distribute the experimental data in a single file or small set of files containing all the parameters and results ordered properly in time. Second, due to its foundational significance, interest in the results of the experiment is likely quite high, and it should be a goal to distribute the data for independent analysis in a way that is not specific to a particular toolset or environment. MATLAB is an expensive program that may be an obstacle for some researchers. MATLAB is also slow and performing repeated data analyses over the required several degrees of freedom, or searching parameter spaces (such as the delay offsets) as described earlier is not practically feasible. Conversely, with a simple text format for the raw data, for example, a C program could be used to load and efficiently analyze the data (C compilers are ubiquitous). Third, the data is stored in descending time order within each MATLAB file while the file numbering is ascending. This again complicates data extraction and we will see that analysis is more convenient with globally ascending timetags. Finally, the data as published is missing half of the Pockels cell opening events and the authors specify that these events must be added back during analysis. It is better to do that in the common distributed data than to require every analyst to make this correction.

For the reasons cited, I extract the data from the authors' data distribution using the procedure described below to produce a set of twenty text files named 'data1.txt' through 'data20.txt' corresponding to the twenty directories of the published data distribution. Each text file contains events, one per line, for a period of time, represented in the following way:

```
3937439023548 12 15
3937439279548 12 15
3937439535574 12 15
3937439791574 12 15
3937440047599 12 15
3937440303599 12 15
3937440559624 12 15
[etc.]
```

The first field of each line is the extracted timetag of the event in the original time unit of 156.25 picoseconds, as distributed in the raw data. The second field represents the angle settings for the event (11, 12, 21, or 22, as described in the authors' data notes). The third field is the detection event type: value 15 denotes an opening of the Pockels cell; value 1 denotes a photon detection at side 1; and value 2 denotes a photon detection at side 2.

The extracted text files are created by a two-step process applied to each of the twenty directories of the data distribution. For example, the file 'data1.txt' is created by first running the MATLAB program 'Extract.m' [6] and then sorting the result text file in ascending order of timestamps. It is important to realize that *this data extraction is lossless* in the sense that the original events are not edited or selected in any way. The missing Pockels cell opening events are added; the data is sorted in ascending order, and the data is coded in text files. I have made these extracted text data files available to researchers interested in performing an independent data analysis [6].

It would be possible to concatenate the twenty text data files into a single text file, but this can cause problems for text editors/viewers, which in many cases cannot open such large files. Instead of concatenating the text files, it is useful to *compile* the text files into a single binary file that is much smaller and can be efficiently loaded by analysis software. The next section describes this compilation process.

## 5.2  Data compilation

The extracted data files generated in the previous data extraction step are suitable for dissemination and perusal with text editors and utilities, but they are inconvenient for repeated programmatic analysis using different degrees of freedom for aspects of the analysis (such as using different delay corrections, window sizes, etc.) The extracted data is contained in multiple large text files and, consequently, the load time for the data before analysis can begin is long. To mitigate this, we *compile* the extracted data into a single binary file that can be loaded very quickly.

The compilation process includes two aspects. First, we compress the data by eliminating the very large number of entries for Pockels cell openings containing no detection events. Instead, we store only entries for each detection event (in ascending time order) and keep counts of the openings (refer to the binary structure description below). We also compress by storing binary values rather than text-encoded values. As a consequence of the compression we can conveniently place all the detection events into a single binary file. Second, we convert the original time values from 156.25 picosecond granularity to 1 picosecond granularity.

The binary structure of the compiled data file is described by the following C code:

```
#define a1b1 0
#define a1b2 1
#define a2b1 2
#define a2b2 3

U32 num_detection_events;
U32 total_trials[4];

struct EVENT
{
    double raw_time;
    double pockels_time;
    unsigned char angles;
    unsigned char channel;
    unsigned int trials[4];
} events[num_detection_events];
```

The `total_trials` field contains the total number of Pockels cell openings (trials) for each of the angle combinations. These counts are used when performing an overall analysis of the data (without partitioning). The `raw_time` field is the detection time of the event in picoseconds and represented as a double floating point value. The `pockels_time` field is the time of the Pockels cell opening just preceding the raw time of the detection event. During analysis we allow for the possibility that a time-of-flight-adjusted detection time may move the event into the previous opening cycle. The `angles` field specifies the angle settings in effect at the time of the detection event. For example, an angle specification of '11' in the extracted text data files corresponds to the definition `a1b1`. The `channel` field contains the value 1 for a detection event at side 1 and 2 for a detection event at side 2. The `trials` array field contains counts, one per angle setting combination, of the number of Pockels cell openings, up to and including, the current detection event. These counts allow for a partition-based analysis of the data.

Again, it is important to realize that *data compilation is lossless* in the sense that the original events are not edited or selected in any way. The resulting binary data file is used in the final phase, the actual analysis of the data to generate the CH metrics. The C language program that performs the compilation and the resulting compiled data file are available online [6].

### 5.3   Data analysis

The computer program used to analyze the experimental data and calculate the CH metrics is available online [6]. The partition size, no averaging/averaging, delay set, and window size are selectable. The events are read into memory from the compiled binary experimental data file. They are adjusted with the required corrections for the selected delay set. The partitions are equal-sized contiguous subsets of the event data. The CH metrics are calculated and displayed. The pairing algorithm operates as follows:

```
for each partition
    initialize counts to 0
    for each trial in the partition that has detection events
      count side 1 detections in that trial within the opening window and add to side 1 singles total
      count side 2 detections in that trial within the opening window and add to side 2 singles total
      add min(side 1 count, side 2 count) to the coincidences total
    calculate and record partition metrics
```

Table 4 shows the CH positivity for the experimental data when analyzed with a partition size of 10,000, with and without singles-rate averaging, and with either the Christensen or CH-optimal delay sets. It can be seen that the chosen delay set has a large influence on the resulting positivity, and this underscores the need to have a correctly designed and implemented event pairing algorithm. It would be possible to make a small mistake in the design or implementation of the pairing algorithm, perhaps inadvertent, or perhaps tied to some extra hypothesis we (falsely) think is true, such that the analysis falsely shows CH violation, or conversely, falsely shows no violation. For these reasons, the pairing algorithm used here is fully described and the implementation source code is available [6] and can be checked for correctness.

We also see that singles-rate averaging diminishes the positivity, just as it did for the quantum joint prediction. Even with the most favorable analysis (no averaging and CH-optimal delays), the positivity is well below 0.5 (any statistically significant positivity greater than 0.5 indicates a violation of the CH inequality). The corresponding quantum joint prediction for the corresponding conditions is 0.847, well above 0.5.

**Table 4. Positivity for the Christensen *et al.* experimental data for several analytical degrees of freedom (partition size 10,000, window size 2.5 μsec)**

| Without singles-rate averaging | | With singles-rate averaging | |
|---|---|---|---|
| Christensen delay set (2.65 μsec/ 2.525 μsec) | CH-optimal delay set (1.292 μsec/ 1.195 μsec) | Christensen delay set (2.65 μsec/ 2.525 μsec) | CH-optimal delay set (1.292 μsec/ 1.195 μsec) |
| 0.178947 | 0.494737 | 0.073684 | 0.452632 |

Continuing with the conditions of analysis most favorable to finding CH violations, Table 5 shows the positivity at various partition sizes. If we choose very small partitions, not only will the statistics be less reliable, but there is the possibility that one or more partitions may not have samples from all four measurement angle sets, and these partitions are insufficient and unusable. The third column in Table 5 shows the number of insufficient partitions out of the total number of partitions. With partition sizes of around 10,000 and above, the insufficient partitions become insignificant. Note that the absolute CH metric corresponding to the last row of Table 5 (one partition) is -0.000032, which does not violate CH.

We are interested in positivities greater than 0.5, and in Table 5 we see three of them, at partition sizes 5,000, 50,000, and 400,000. This might at first seem favorable to the quantum joint prediction. However, all of these cases of apparent positivity greater than 0.5 are statistically insignificant. For example, in the case of partition size 50,000, a binomial random distribution gives an expected value of 10 positive partitions out of 20, with standard deviation of 2.236. In this case there were 11 positive partitions out of 20. That is well below one standard deviation different from the expected value. There are no statistically significant violations of the 50% rule for any of the partition sizes in Table 5.

Of particular interest in Table 5 is the case of partition size 1500, with 652 partitions. Christensen *et al.* claimed 394 positive partitions out of 650, which is statistically significant to more than 5 standard deviations. However, for the corresponding case in the present analysis of the experimental data, I find the positivity to be only 0.452632 (172 out of 380 sufficient partitions [recall, a sufficient partition contains events representing all 4 possible measurement angle setting combinations]). As it is not possible to find 650 equal-sized contiguous sufficient partitions in the published data, the Christensen *et al.* claim of 394 out of 650 must be questioned. Perhaps they have included additional unpublished data in their analysis, or they have chosen partitions in some other non-straightforward way. Any special post-selection of partitions must be viewed with suspicion.

Very striking also in Table 5 is the very different behavior of the metric as the partition size is varied. In Table 2, the positivity for the quantum joint prediction starts around 0.5 at a small partition size and smoothly converges to positivity 1.0 as the partition size is increased. However, in Table 5, the experimental positivity goes to 0.0 at the largest partition size and there is no statistically significant positivity at any other partition size.

 

**Table 5. Experimental positivity at various partition sizes**
**(without singles-rate averaging, with CH-optimal delay set, window size 2.5 μsec)**

| Partition size | Positivity | Insufficient partitions |
|---|---|---|
| 100 | Insufficient data | all |
| 500 | 0.483871 | 1861 of 1954 |
| 1000 | 0.458824 | 637 of 977 |
| 1500 | 0.452632 | 272 of 652 |
| 5000 | 0.522222 | 16 of 196 |
| 10,000 | 0.494737 | 3 of 98 |
| 25,000 | 0.461538 | 1 of 40 |
| 50,000 | 0.550000 | 0 of 20 |
| 75,000 | 0.384615 | 1 of 14 |
| 100,000 | 0.500000 | 0 of 10 |
| 150,000 | 0.428571 | 0 of 7 |
| 200,000 | 0.400000 | 0 of 5 |
| 300,000 | 0.500000 | 0 of 4 |
| 400,000 | 0.666667 | 0 of 3 |
| 500,000 | 0.500000 | 0 of 2 |
| [one partition] | 0.000000 | 0 of 1 |

**Table 6. Experimental positivity at various window sizes**
**(without singles-rate averaging, with CH-optimal delay set, partition size 10,000)**

| Window size (μsec) | Positivity |
|---|---|
| 0.50 | 0.094737 |
| 0.75 | 0.231579 |
| 1.00 | 0.347368 |
| 1.25 | 0.400000 |
| 1.50 | 0.389474 |
| 1.75 | 0.421053 |
| 2.00 | 0.431579 |
| 2.25 | 0.389474 |
| 2.50 | 0.494737 |
| 2.75 | 0.463158 |
| 3.00 | 0.442105 |

## 5.4    Results of the data analysis

The foregoing analysis of the Christensen *et al*. experimental data shows no statistically significant violation of the CH inequality, and behavior that clearly does not match the quantum joint prediction. This disconfirms the applicability of the quantum joint prediction, and excludes nonlocal information sharing between the sides. The experimental data provides evidence that confirms local realism. The positivity is never significantly greater than 0.5 and it converges to 0.0 at the largest partition size (including all of the data in one partition). This agrees with our intuition, because we know that a joint PDF cannot be sampled by separated (marginal) measurements, and that special relativity entails locality [2].

# 6.  ALTERNATIVE ANALYSIS CONSIDERING ACCIDENTAL COINCIDENCES

An objection to the foregoing analysis could proceed as follows. The coincidence window size of 2500 nanoseconds used in the analysis is large, compared for example to the detection jitter (a few hundreds of nanoseconds). Such a large window may include a large number of accidental coincidences, the inclusion of which reduces the CH metric, converting a violation of CH into a non-violation. It is important to dispel this objection by showing that accidental coincidences are negligible in the Christensen *et al*. experiment.

An accidental coincidence is tallied when two source pair events are emitted in the time span of the coincidence window, and for the first pair, one side detects it while the other does not, and vice versa for the second pair. The two resulting singles are erroneously counted as a coincidence. At first glance it might seem that extra coincidences in all the experiments would *increase* the CH metric, because three of the coincidence terms in the inequality are positive and only one is negative. However, the measurement angles for the experiment ensure that the accidentals are much greater in the a2b2 experiment, which is weighted negatively in the CH inequality, than the other three experiments (because the product of the singles is much greater in the a2b2 experiment). Thus accidental coincidences can decrease the CH metric.

Based on known theoretical models of coincidence counting, it can be easily shown that the derivative C' = dC/dW of the total number of coincidences with respect to the window size in an experiment is a function of the square of the source pair production rate R. In an experiment where the window size is larger than the detection jitter (to avoid losing true coincidences), if no accidentals are present then C' will be 0, because after all the true coincidences are tallied further increases in the window size do not tally any accidentals. This leads us to the conclusion that for a valid experiment, we must reduce the source pair production rate relative to the window size sufficiently to ensure that accidental coincidences are negligible. Clauser and Horne [1] long ago warned experimentalists about this requirement:

> *Thus, we tacitly require the experimental arrangement to be such that this condition obtains (suitable source strength, time separation of pairs, etc.). If a sufficiently weak source is used, the ratio of "chance" coincident counts to "true" coincident counts can be made arbitrarily small, and the corresponding dead time can also be minimized.*

To assess the role of accidental coincidences in the Christensen *et al*. experiment I developed an alternative analysis using a variable coincidence window size. First, the detection events were corrected for time-of-flight as previously described. Next, detection events outside a Pockels cell opening were removed (they are noise). Then a list of remaining events was created and sorted for each experiment. These lists were analyzed for coincidences using a conventional coincidence counting algorithm based on a 'greedy' method with a variable window size. The results are shown in Figure 1. It can be seen that at a window size greater than about 500 nanoseconds, all the true coincidences have been tallied and that the slopes (C') become very close to 0. This proves that accidentals are negligible in the Christensen *et al*. experiment. Also note that the CH metric never exceeds 0, again confirming local realism.

When accidental coincidences are negligible, the full-counting method used in the main analysis of this paper is fully justified. My analyses show that full counting, the greedy analysis, and the Giustina *et al*.-style analysis [7] all produce very similar coincidence counts. Any one of these is adequate and justified when accidental coincidences are negligible, although one might prefer the full-counting method due to its intuitive nature and utter simplicity of implementation.

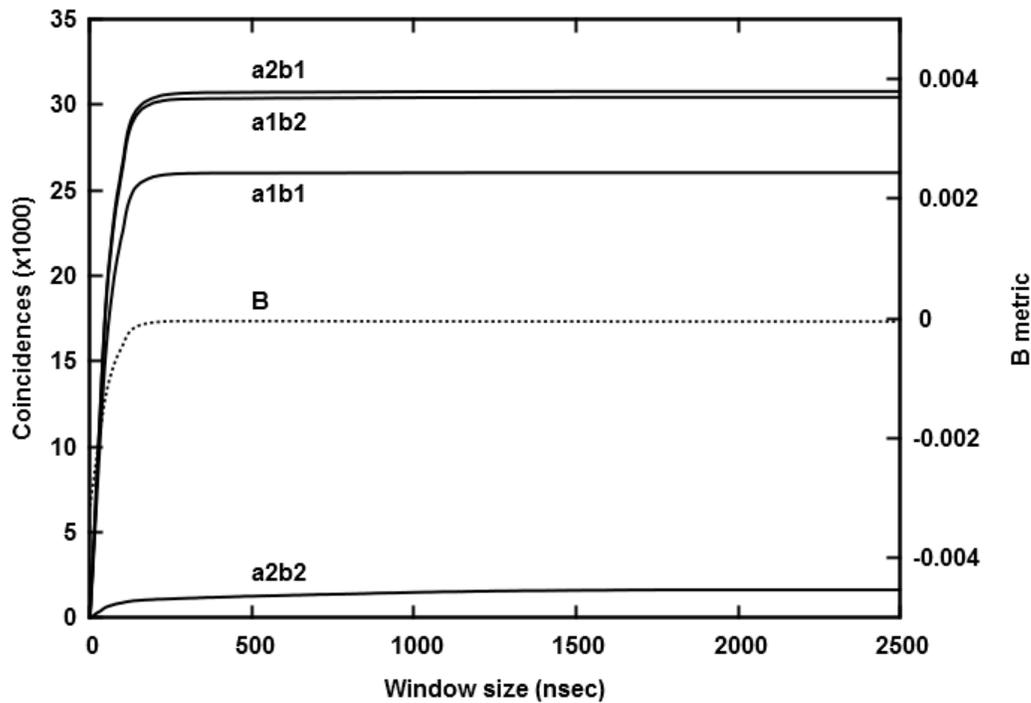

**Figure 1. Coincidence counts and CH metric versus coincidence window size in the Christensen *et al*. experiment**

By way of contrast I also show in Figure 2 a similar plot for the Giustina *et al*. experiment [8]. It can be seen that accidental coincidences are not negligible, and are especially large in the a2b2 experiment. The experiment fails miserably in satisfying Clauser and Horne's stipulation about proper experimental design. There are a number of other deficiencies in the Giustina *et al*. experiment and future work will attempt to account for the small negative dip of J at a window size of about 100 nanoseconds. One possible mechanism relies on differing jitter distributions for the four experiments. The Christensen *et al*. experiment avoids these deficiencies and, as accidentals have been shown to be negligible, the objection to full counting based on accidental coincidences is dispelled.

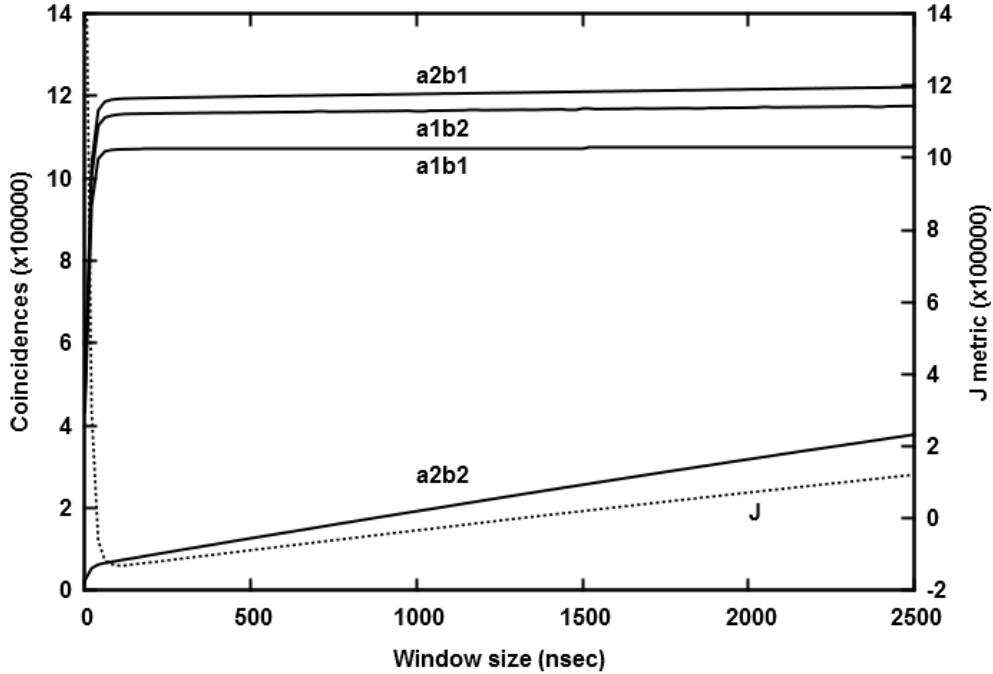

**Figure 2. Coincidence counts and Eberhard metric versus coincidence window size in the Giustina *et al*. experiment**

# 7. CONCLUSION

Christensen *et al*. present us with an experimental refutation of quantum nonlocality. The experimental data *confirms local realism* and *disconfirms quantum nonlocality* (i.e., the spooky sampling of a joint PDF by means of separated [marginal] measurements). The Christensen *et al*. experiment is decisive because it is fully adequate in that the detection efficiency is high enough, the noise is low enough, and the accidental coincidences are low enough to show a violation of the CH inequality, if it were present. In the future, it will be exciting to analyze the data for other recent high-efficiency EPRB experiments [8] to perform independent data analyses like the one in this paper. Future work should also include development of simulations that show how local mechanisms can produce the artifactual violations reported here. A simulation must produce statistics like those of the Christensen *et al*. experiment, including the contradictory results of the two counting methods, i.e., including all the events in a timeslot does not violate the inequality, while picking just the first one per timeslot does violate it.

As the current interpretations of the EPR paradox leave us unsatisfied, and the experimental results described force a revision of current thinking, I offer an executive summary of what one may call a 'rational interpretation' of EPR:

*a.* The quantum joint prediction cannot be recovered in an experiment with separated (marginal) measurements, just as for classical probability. Quantum mechanics correctly applied does not predict a violation of the CH inequality. The correct quantum mechanics prediction for an EPRB experiment must use the marginals (via reduced density matrices) and not the joint distribution. The source distribution in an EPRB experiment may be a joint one, but joint statistics cannot be recovered because the experiment yields only separated (marginal) measurements. A well-developed statistical field of study decomposes correlated joint distributions into the marginals plus an additional function called a *copula*. There would be no need for this field if any arbitrary joint distribution could be recovered through its marginals. Therefore, we cannot and do not apply the quantum joint prediction to EPRB experiments. As this point has appeared obscure to some, I remark further as follows: The joint distribution is fixed by the physics of the photon-pair source generating the singlet stream. The question is whether that distribution can be recovered (sampled) from separated (marginal) measurements. I have shown elsewhere [2] that the source distribution cannot be recovered (sampled) in an EPRB experiment. One might think of the post-selection (data discarding) that we have seen in the Christensen *et al*. analysis as a sort of perverse copula, inserted prior to the correlation of the outcome streams. The result, however, is arbitrary and artificial; the desired result is simply engineered through the *ad hoc* perverse copula. The quantum joint prediction predicts a violation even for direct correlation without any copula (the rational way experiments were analyzed before nonlocalists realized that CH could not be violated that way), so why would we think of inserting an arbitrary, unjustified copula? There may be some circumstances or experiments where a joint distribution can be jointly sampled and recovered [2], but an EPRB experiment is not one of them. When this delusion about joint sampling from marginals, inexplicably accepted without question, is overcome everything falls into place.

b.  Quantum mechanics is not therefore wrong or disproved. The essence of quantum mechanics is completely unaffected; we need only to be careful about separated measurement situations, just as we are in classical probability theory. Just as we would not blindly expect the quantum joint prediction to apply in the presence of heavy decoherence, we should not expect it to apply in a case of separated (marginal) measurement.

c.  Valid experiments properly interpreted, such as Christensen *et al*., do not violate the CH inequality, and therefore confirm local realism. In an EPRB experiment all the experimental events must be included on an equal basis. Methods using post-selection, such as the method used by Christensen *et al*., produce only artifactual violations of CH due to systematic data discarding.

d.  John Bell's work is not challenged, with one caveat: even quantum theory must face the no-go results. It is only the misguided idea that a joint distribution can be sampled with marginal measurements that led to the mistake of thinking that quantum mechanics predicts a violation.

e.  This 'rational interpretation' resolves the EPR paradox. It has always been the irrational idea of quantum nonlocality that blocked proper understanding. Without nonlocality, a consistent axiom set for physics is restored, along with our intuitions.

The 'rational interpretation' completely turns the table on the quantum nonlocalists. A local realist who asserts that quantum mechanics does not predict, nor do experiments manifest, violation of the CH inequality is relieved of any need for 'loopholes' purported to allow for classical violations. Put dramatically, a quantum nonlocalist asks "you've found another loophole?", and the local realist counters "no, I have exposed yours". If a local realist claims and reports no violation, then loopholes can arise only on the quantum nonlocalist side, embodied in such sleight-of-hand as post-selection. Local realists simply assert that valid experiments cannot and do not show CH violation, and locality is confirmed. Meanwhile, prominent quantum nonlocalists continue to brazenly assert that nonlocality cannot be experimentally disconfirmed – science has become faith. Never doubt that passion flies at a distance!

Raw nature will brutally crush the delusion; quantum nonlocality will come to be seen as the modern analog of perpetual motion. Rational quantum theorists accept that the quantum joint prediction cannot be obtained from separated (marginal) measurements in an EPRB experiment, and that information cannot be nonlocally transferred in violation of special relativity. Such theorists therefore expect all valid experiments to confirm local realism and expect the quantum joint prediction to be disconfirmed in experiments. It is no surprise that the Christensen *et al*. experiment decisively confirms local reality.

## Acknowledgements

The author is grateful to the Christensen *et al*. team for making available the raw data from their experiment, and to Brad Christensen personally for extensive and highly useful discussions, and for his making available the MATLAB code that implements the data analysis used by Christensen *et al*. to generate their published results.

## References

[1]  Clauser, J. F., and Horne, M. A., "Experimental consequences of objective local theories", *Phys. Rev. D.* 10, No. 2, 526-535 (1974).

[2]  Graft, D.A., "On Bell-Like Inequalities for Testing Local Realism", arXiv: quant-ph 1404.4329 (2014).

[3]  Christensen, B.G., McCusker, K.T., Altepeter, J.B., Calkins, B., Lim, C.C.W., Gisin, N., and Kwiat, P.G., "Detection-Loophole-Free Test of Quantum Nonlocality, and Applications", *Phys. Rev. Lett*. 111, 130406 (2013).

[4]  The raw data from the Christensen *et al*. experiment is available at the following URL: http://research.physics.illinois.edu/QI/Photonics/research/#tests-of-nonlocality.

[5]  Eberhard, P. H., "Background level and counter efficiencies required for a loophole-free Einstein-Podolsky-Rosen experiment", *Phys. Rev. A*, Vol. 47, No. 2, R747-R750 (1993).

[6]  The extracted and compiled data and all program source code are available at the following URL: http://rationalqm.us/papers/Christensen/.

[7]  Giustina, M., private communication.

[8]  Giustina, M., Mech, A., Ramelow, S., Wittmann, B., Kofler, J., Lita, A., Calkins, B., Gerrits, T., Sae Woo Nam, Ursin, R., and Zeilinger, A., "Bell violation with entangled photons, free of the fair-sampling assumption", *Nature* 497, 227 (2013).